\documentstyle[12pt]{article}

\begin{document}

\title{The 5 March 1979 soft gamma-ray repeater: a fresh outlook }

\author{G.V. Vlasov \thanks{%
E-mail: vs@itp.ac.ru}}

\maketitle

\begin{abstract}
An attempt is made to explain the 8-second periodicity of the 5th March
1979
soft gamma-ray repeater as a consequence of the elastic wave precession
in
the rigid crust of the rotating neutron star.
\end{abstract}

\sloppy

The soft gamma-repeaters (SGR) as well as the ordinary gamma-ray
bursts~(GRB) are observed for almost thirty years. While the origin of
GRB
still remains mysterious~\cite{Kouveliotou+94}, the SGR are associated
with
the neutron
stars~\cite{KF93,Kouveliotou+94,Murakami+94,Kulkarni+94,CD98},
particularly, the most incomparable event enrolled on the 5th of March
1979 
\cite{Mazets+79} from SGR~0526-66, which arouse most inquiry and
discussion.
Its source is believed to be a neutron star, the remainder of
supernova~N~49~%
\cite{HL79,Evans+80,Cline+82,Kouveliotou+94}. The radiation of that SGR
is
modulated at definite frequencies~\cite{Barat+79,Terrel+80}, whereas the
most intense 8-second peak is usually associated with period $P$ of star
rotation~\cite{Kazanas88,AFO86}, the origin of the 4-second \cite
{Barat+79,Terrel+80} and 6-second~peaks \cite{Barat+79} remains unknown.
The
rigid crust radial vibration is suggested to take place in such
events~\cite
{Ramaty+80,Kazanas88}, yet, those periodicities cannot be provoked by
vibration itself, since the relevant characteristic frequencies ($\nu
\sim
1ms^{-1}$) are too high (other examples of SGR with periods
$P\symbol{126}1s$
are also known \cite{Kouveliotou+98}) . We guess the nature of those
''attendant'' frequencies, as well as the main period, can be explained
as
soon as we appeal to an interesting phenomenon neglected so far. It is
the
inertia of elastic waves in a rotating rigid body, first discovered a
century ago \cite{Bryan1890}. The investigation of a new gyroscope \cite
{Scott82,Zhuravlev93} was nourished from this effect.

Suppose that the elastic vibrations are excited in the solid crust of a
neutron star. The spectrum of vibrations is a discrete set of
frequencies $%
\nu _k$ denoted by integer $k$. If the body is put into rotation at
frequency $\omega =2\pi /P$, then, for each $k$-mode, there exists a
reference frame within which the relevant standing wave is possible. But
this reference frame rotates itself at definite angular velocity \cite
{Zhuravlev93} 
\begin{equation}
\Omega =\omega \left( 1-\chi \right) \qquad \qquad \chi =\frac{kA_n^2}{%
1+n\left( n+1\right) A_n^2}\qquad 2\leq k\leq n\qquad 0<\chi <1 
\label{o}
\end{equation}
where parameter $A_n$ is a function of the number of waves $n$ along the
meridian and the Poisson coefficient $\mu $ \cite{Zhuravlev86}:

\begin{equation}
A_n=\frac{-2\left( 1+\mu \right) +\frac 12\left( 1-\mu \right) \lambda
}{%
\left( 1+\mu \right) n\left( n+1\right) }  \label{a}
\end{equation}
where $\lambda $ is determined by equation 
\begin{equation}
\left( 1-\mu \right) \lambda ^2-2\nu \left[ 1+3\mu +n\left( n+1\right)
\right] +4\left( 1+\mu \right) \left( n^2+n-2\right) =0  \label{nu}
\end{equation}
By the way, the auxiliary modes (marked by tilde) are also possible
here.
For the body at rest all $\Omega $ are the same and equal to $\omega $.
In
other words, one finds the vibration modes traveling, each at definite
period $T=2\pi /\Omega $ and not equal to $P$. In addition, the larger
$k$,
the lesser $T$, so that $\Omega \rightarrow \omega $ when $k\rightarrow
\infty $. This phenomenon is known as inertia of the elastic waves, for
the
elastic vibrations behave as if they were a mass involved into
precession.

For the matter of the neutron star crust \cite{Ruderman91a}~with $\mu
\simeq
1/2$ the main and auxiliary period of the simplest dipole mode
($n=k=2$),
obtained in the frames of theory~\cite{Zhuravlev93} by Eqs.
(\ref{o})-(\ref
{nu}), is estimated as $1.28P$ and $1.13P$ respectively.

However, for $n=2$ and $k=4$ we come to a system of fractional
connectedness. In the light of crustal plate tectonics during
gamma-bursts 
\cite{Ruderman91a,Ruderman91b,LFE98}, the crust cracking and splitting
in
two halves is the simplest event of conceivable situations. Indeed, the
case 
$k=2n$, corresponds to the two vibrating semi-spheres. The formula
(\ref{o})
then is reduced to \cite{ZP85} 
\begin{equation}
\chi =\frac{-2-\mu \pm \sqrt{\left( 2+\mu \right) ^2+4\left( 1-\mu
\right)
k^2}}{2\left( 1-\mu \right) k^2}\simeq \frac{-5\pm
\sqrt{25+8k^2}}{2k^2}%
\qquad  \label{o2}
\end{equation}
and it gives $T=1.30P$ and $\tilde T=0.65P$ at $k=4$; if $P=6s$, then $%
T=7.8s $ and $\tilde T=3.9s$. Of course, we may vary $k$ to obtain
various
ratio $T/\tilde T$.

If we rely on the theory of relationship between gamma-radiation and
star
crust vibration, as it has been claimed once~\cite{Ramaty+80}, the
observer
will reveal radiation modulated at frequencies $\Omega $ determined by
the
formulae (\ref{o}), (\ref{o2}) and in agreement with the observations
\cite
{Barat+79}. Note that the frequency $\omega $ itself may not be
discovered
at all. Were there no inertia of elastic waves in a rigid crust, the
observer would detect only the single frequency $\omega $ instead of a
set $%
\left\{ \Omega \right\} $.

Whereas, the real process includes other factors not discussed in this
simple consideration, it is clear that the most intensive 8-second peak
scarcely coincides with the period of star rotation. Since the lowest
modes
are the most powerful, the 8-second period probably corresponds to the
main
dipole mode of a vibrating sphere or two semi-spheres. Although we have
not
taken into account all the details, for instance the general
relativistic e
ffects, the role of finite ratio $\Omega /\nu _k$ and thickness of the
crust, the qualitative result remains the same anyhow. The precession of
elastic vibrations in the solid crust of a rotating neutron star leads
to a
qualitatively new effect and the radiation of SGR shall be modulated at
several frequencies instead of a single one.


\begin{thebibliography}{99}
\bibitem{Kouveliotou+94}  Ch. Kouveliotou {\it et al}, Nature {\bf 368},
125
(1994).

\bibitem{KF93}  S.R. Kulkarni and D.A. Frail, Nature {\bf 365}, 33
(1993).

\bibitem{Murakami+94}  T. Murakami {\it et al}, Nature {\bf 368}, 127
(1994).

\bibitem{Kulkarni+94}  S.R. Kulkarni {\it et al}, Nature {\bf 368}, 129
(1994).

\bibitem{CD98}  K.S. Cheng and Z.G.Dai, Phys. Rev. Lett. {\bf 80}, 18
(1998).

\bibitem{Mazets+79}  E.P. Mazets {\it et al}, Nature {\bf 282}, 587
(1979).

\bibitem{HL79}  D.J. Helfand and K.S. Long, Nature {\bf 282}, 589
(1979).

\bibitem{Evans+80}  W.D. Evans {\it et al}, Astrophys. J. {\bf 237}, L7
(1980).

\bibitem{Cline+82}  T.L. Cline {\it et al}, Astrophys. J. {\bf 255}, L45
(1982).

\bibitem{Barat+79}  C. Barat {\it et al}, Astron. Astrophys. {\bf 79},
L24
(1979).

\bibitem{Terrel+80}  J. Terrell {\it et al}, Nature {\bf 285}, 383
(1980).

\bibitem{Kazanas88}  D. Kazanas, Nature {\bf 331}, 320 (1988).

\bibitem{AFO86}  Ch. Alcock, E. Farthi, and A. Olinto, Phys. Rev. Lett.
{\bf %
57}, 2088 (1986).

\bibitem{Ramaty+80}  R. Ramaty {\it et al}, Nature {\bf 287}, 122
(1980).

\bibitem{Kouveliotou+98}  Ch. Kouveliotou {\it et al}, astro-ph/9809140.

\bibitem{Bryan1890}  G.H. Bryan, Proc. Cam. Philos. Soc. Math. Phys.
Sci. 
{\bf 7}, 101 (1890).

\bibitem{Scott82}  W.B. Scott, Aviat. Week \& Space Technol. {\bf 117},
n.17, 64 (1982).

\bibitem{Zhuravlev93}  V.F. Zhuravlev, Mech. Solids {\bf 28}, n.3, 3
(1993).

\bibitem{Zhuravlev86}  V.F. Zhuravlev, Mech. Solids {\bf 21}, n.6, 87
(1986).

\bibitem{Ruderman91a}  M. Ruderman, Astrophys. J. {\bf 366}, 261 (1991).

\bibitem{Ruderman91b}  M. Ruderman, Astrophys. J. {\bf 382}, 587 (1991).

\bibitem{LFE98}  B. Link, L.M. Franco and R.I. Epstein,
astro-ph/9805115.

\bibitem{ZP85}  V.F. Zhuravlev and A.L. Popov, Mech. Solids {\bf 20}
n.1,
138 (1985).

\end{thebibliography}
\end{document}